\documentclass[graybox]{svmult}
\usepackage[colorlinks=true]{hyperref}

\newcommand{\comment}[1]{}

\newcommand{\lr}[1]{ \left( #1 \right) }

\newcommand{\vev}[1]{ \langle \, #1 \, \rangle }

\usepackage{mathptmx}       
\usepackage{helvet}         
\usepackage{courier}        
\usepackage{type1cm}        
\usepackage{makeidx}         
\usepackage{graphicx}        
\usepackage{multicol}        
\usepackage[bottom]{footmisc}

\makeindex             

\begin{document}

\title*{Lattice studies of magnetic phenomena in heavy-ion collisions}
\author{P. V. Buividovich, M. I. Polikarpov and O. V. Teryaev}
\institute{P. V. Buividovich \at Institute of Theoretical Physics, University of Regensburg, D-93053 Germany, Regensburg, Universit\"{a}tsstra{\ss}e 31, \email{Pavel.Buividovich@physik.uni-regensburg.de}
\and M. I. Polikarpov \at ITEP, 117218 Russia, Moscow, B. Cheremushkinskaya str. 25, \email{polykarp@itep.ru}
\and O. V. Teryaev \at JINR, 141980 Russia, Dubna, Joliot-Curie str. 6,
\email{teryaev@theor.jinr.ru} }
%
%
\maketitle

\abstract{We review some experimental consequences of the presence of superstrong magnetic fields of order of the nuclear scale in noncentral heavy-ion collisions. We present lattice estimates for the strength of the Chiral Magnetic Effect (CME) for different quark flavours and argue that the dependence of the anisotropy of the distribution of emitted hadrons on their flavor content might be used as another experimental evidence of the CME. Another possible effect of superstrong magnetic field might be the observed abnormal enhancement of dilepton yield. We show that the presence of the magnetic field leads to a specific anisotropy of the dilepton emission rate.}

\section{Introduction}
\label{sec:introduction}

 It has been realized recently that in heavy-ion collision experiments hadronic matter is affected not only by extremely high temperatures and densities, but also by superstrong magnetic fields with field strength being comparable to hadron masses squared. Such superstrong fields are created due to the relative motion of heavy ions themselves, since they carry large charge $Z \sim 100$ \cite{Kharzeev:08:1}.

 Obviously, the magnetic field is perpendicular to the collision plane, which can be reconstructed in experiment from the angular distribution of produced hadrons \cite{Abelev:09:1}. There is no direct experimental way to measure the absolute value of the field strength, but it can be estimated in some microscopic transport model, such as the Ultrarelativistic Quantum Molecular Dynamics model (UrQMD) \cite{Skokov:09:1}.

 Probably the most notable effect which arises due to magnetic fields in heavy-ion collisions is the so-called Chiral Magnetic Effect. The essence of the effect is the generation of electric current  along the direction of the external magnetic field in the background of topologically nontrivial gauge field configurations \cite{Kharzeev:08:1, Kharzeev:08:2}. Such generation is not prohibited by $\mathcal{P}$-invariance, since topological charge density is a pseudoscalar field and thus nonzero topological charge explicitly breaks parity. However, since QCD is parity-invariant, the net current or the net electric charge should vanish when averaged over multiple collision events. Nevertheless, the nontrivial effect can still be detected if one considers dispersions of electric current or electric charge \cite{Buividovich:09:7, Shevchenko:10:1}. Experimentally, the Chiral Magnetic Effect manifests itself as the dynamical enhancement of fluctuations of the numbers of charged hadrons emitted above and below the reaction plane \cite{Selyuzhenkov:06:1, Abelev:09:1, Voloshin:04:1, Voloshin:08:1}.

 Another effect, which is closely related to the CME, but has different experimental signatures, is the anisotropic electric conductivity of hadronic matter in the strong magnetic field, which was discovered in lattice simulations \cite{Buividovich:10:1}. Since the conductivity of the hadronic matter is directly related to the lepton emission rate \cite{Teryaev:95:1, Gupta:04:1}, such anisotropic conductivity should result in specific anisotropy of the dilepton emission rate w.r.t. the reaction plane. This anisotropy should grow with the centrality of the collision and with the charges of the colliding ions. This effect might also contribute to the observed abnormal dilepton yield in heavy-ion collisions \cite{Adare:09:1}. Some theoretical considerations \cite{Chernodub:10:1} as well as preliminary lattice data \cite{Buividovich:11:2} suggest that at very strong magnetic fields with strength $e B > m_{\rho}^2$, where $m_{\rho}$ is the mass of the charged $\rho$-meson, the anisotropic conductivity might even turn into the anisotropic superconductivity in the direction of the field. Unfortunately, such extremely large field strengths are hardly reachable with present-day heavy-ion colliders.

 In this paper we give some estimates of the expected experimental signatures of superstrong magnetic fields, basing on the lattice data. In Section \ref{sec:CME} we consider the Chiral Magnetic Effect and argue that its strength should decrease with increasing quark mass, which can be used to discriminate between the CME and other possible effects which might result in preferential emission of charged hadrons in the direction perpendicular to the reaction plane. In Section \ref{sec:dilepton} we consider the dilepton emission rate, and estimate the contribution of the induced conductivity to the total dilepton yield and dilepton angular distribution in heavy-ion collisions.

\section{Chiral Magnetic Effect}
\label{sec:CME}

 Chiral Magnetic Effect is usually characterized by the following experimental observables, suggested first in Ref. \cite{Voloshin:04:1}:
\begin{eqnarray}
\label{observables}
a_{ab} = \frac{1}{N_{e}} \, \sum \limits_{e = 1}^{N_{e}} \, \frac{1}{N_{a} N_{b}} \,
\sum \limits_{i=1}^{N_{a}} \sum \limits_{j=1}^{N_{b}} \, \cos\lr{\phi_{i a} + \phi_{j b}},
\end{eqnarray}
where $a, b = \pm$ denotes hadrons with positive or negative charges, respectively, $N_{e}$ is the number of events used for data analysis, $N_{a}$ and $N_{b}$ are the total multiplicities of positively/negatively charged particles produced in each event, and $\phi_{i a}$, $\phi_{j b}$ are the angles w.r.t. the reaction plane at which the hadrons with labels $i$ and $j$ are emitted. Summation in (\ref{observables}) goes over all produced hadrons. In practice, only sufficiently energetic particles are considered.

 The main signatures of the CME is the growth of $a_{ab}$ with impact parameter, as well as the negativity of $a_{++}$ and $a_{--}$ and the positivity of $a_{+-}$ \cite{Selyuzhenkov:06:1, Abelev:09:1, Voloshin:04:1, Voloshin:08:1}. However, it has been pointed out recently that these results can also be explained by other effects,
such as the influence of nuclear medium on jet formation \cite{Petersen:11:1} and other in-medium effects \cite{Ma:11:1}. It is therefore important to think about more refined experimental tests of the CME. Our main message in this paper is that the dependence of the charge fluctuations on quark mass can be used to discriminate between the CME and other possible phenomena which contribute to the observed asymmetry of charge fluctuations. Indeed, the CME emerges due to fluctuations of quark chirality \cite{Kharzeev:08:1}, which are suppressed when the quark mass is increased.

 The dependence of the observables (\ref{observables}) on the quark mass can be studied if one sums separately over charged mesons with different quark content, e.g. $u\bar{d}$ and $d \bar{u}$ (charged pions), $u\bar{s}$ and $\bar{u} s$ (charged kaons) or $\bar{u} c$, $\bar{c} u$, $\bar{d} c$, $d \bar{c}$ ($D$-mesons). The observables $a_{ab}$ should then decrease with the meson mass. The dependence of $a_{ab}$ on the centrality of the collision should also become weaker. Here we give a rough estimate of this effect basing on the results of lattice simulations.

 Let us first note that the observables $a_{a, b}$ can be expressed in terms of the differences of multiplicities of charged hadrons emitted above and below the reaction plane \cite{Kharzeev:08:1}:
\begin{eqnarray}
\label{observables_vs_delta}
a_{ab} = \frac{c \, \vev{\Delta_{a} \Delta_{b}}}{\vev{N_{a}} \vev{N_{b}}}\,,
\end{eqnarray}
where $a, b = \pm$, $\Delta_{\pm}$ are the differences of the multiplicities of hadrons with positive or negative charges above or below the reaction plane, respectively. The factor $c$ depends on the hydrodynamical evolution of hadronic matter, and is usually close to unity. For multiplicities $\vev{N_{a}} \sim 1000$ one can also neglect   the initial charge of heavy ions $Z \sim 100$ with a good precision, and assume that $\vev{N_{a}} = \vev{N_{b}} = N_{q}$, where $N_q$ is the mean multiplicity of the same-charge hadrons per event.

 Lattice results can be compared to experimental data by considering the quantity $a_{++} + a_{--} - 2 a_{+-}$, which can be expressed solely in terms of the difference $\Delta Q = \Delta_{+} - \Delta_{-}$ of net charges of hadrons emitted above and below the reaction plane:
\begin{eqnarray}
\label{main_observable}
 a_{++} + a_{--} - 2 a_{+-} = \frac{\vev{\lr{\Delta Q}^{2}}}{N_{q}^{2}} = \frac{\vev{\lr{\Delta_{+} - \Delta_{-}}^{2}}}{N_{q}^{2}}\, ,
\end{eqnarray}
In turn, the dispersion of the charge difference $\vev{\lr{\Delta Q}^{2}}$ can be related to the vacuum expectation values of the squared current densities $\vev{j_{\mu}^{2}\lr{x}}$ \cite{Buividovich:09:7}. The contribution of each quark flavor $f = u, d, s, c$ to the total electromagnetic current is $j^{f}_{\mu}\lr{x} = \bar{q}^f \gamma_{\mu} q^f$. We do not consider here the third-generation quarks, which are extremely rarely produced in heavy-ion collisions.

\begin{figure}[b]
\sidecaption
\includegraphics[width=6cm]{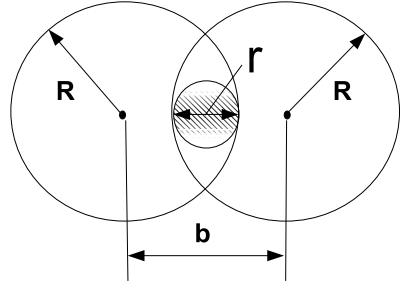}
 \caption{Schematic view of the collision geometry. The fireball is the hatched region of volume $V \sim \frac{4 \pi}{3} \, \lr{R - b/2}^3$ within the intersection of two heavy ions of radius $R$ each.}
  \label{fig:fireball}
\end{figure}

 The simplest model which allows to express $\vev{\lr{\Delta Q}^{2}}$ in terms of $\vev{j_{\mu}^{f \, 2}\lr{x}}$ is the model of spherical fireball, which emits positively and negatively charged hadrons from its surface with intensity proportional to $\vev{j_{\mu}^{f \, 2}\lr{x}}$. This leads to the following relation \cite{Buividovich:09:7}:
\begin{eqnarray}
\label{dQ_curr2_fin}
a_{++} + a_{--} - 2 a_{+-} = \frac{4 \pi \tau^{2} \rho^{2} r^{2}}{3 N_{q}^{2}}\, \lr{\vev{\lr{j^{f}_{\parallel}}^2} + 2 \vev{\lr{j^{f}_{\perp}}^{2}}}  ,
\end{eqnarray}
where $j^{f}_{\parallel}\lr{x}$ and $j^{f}_{\perp}\lr{x}$ are, respectively, the currents along the magnetic field and perpendicular to it, $\tau$ is some characteristic collision time, $r$ is the fireball radius and $\rho$ is some typical correlation length for electric charge density in the fireball. In our estimates, we take $\tau \sim 0.3 \, \mathrm{fm}$ (this is a typical decay time for the magnetic field in heavy-ion collisions \cite{Skokov:09:1}), $\rho \sim 0.2 \, \mathrm{fm}$, which are reasonable parameters for, say, gold-gold collisions at $60 \, GeV /$nuclon. We also assume that the fireball is a sphere with radius $r = R - b/2$ within the overlapping region between the two heavy ions of radius $R$ which collide at impact parameter $b$. The net multiplicity $N_{q}$ and the impact parameter $b$ as the functions of collision centrality can be found in Table 1 in \cite{Kharzeev:08:1}. For simplicity, we assume that $\vev{\lr{j_{\mu}}^{f \, 2}}$ are approximately constant on the surface of the fireball. Note also, that in order to exclude the effects related to the dependence of multiplicities of strange and charmed mesons on the collision centrality, which might be different from that of light mesons, we normalize the charge of emitted hadrons by the square of the total multiplicity of all hadrons.

 Several technical remarks are in order. First, we assume that the matter within the fireball is in the state of thermal equilibrium, and thus the expectation values $\vev{j_{\mu}^{f \, 2}}$ can be calculated from gauge theory in Euclidean space. We also assume that the magnetic field is uniform and nearly time-independent. Of course, these are rather rough approximations, but we are aiming here at qualitative rather than quantitative estimates. We have calculated the currents in $SU(2)$ lattice gauge theory with background magnetic field both in the confinement phase, neglecting the contribution from the virtual quark loops (quenched approximation). A comparison with $SU\lr{3}$ gauge theory suggests that this is a reasonable approximation \cite{Kalaydzhyan:10:1}. A more detailed study has shown also that in the deconfinement phase the dispersions of local current densities are practically independent of the magnetic field \cite{Buividovich:09:7}, thus we do not consider here this case. The expectation values $\vev{j_{\mu}^{f \, 2}}$ contain also the ultraviolet divergent part, which we have removed by subtracting the corresponding expectation values at zero temperature and zero magnetic field.

 The masses of the valence quarks took the values $m_q = 50 \, MeV$ (for $u$ and $d$ quarks), $m_q = 110 \, MeV$ (for $s$-quark) and $m_q = 1 \, GeV$ (for $c$-quark). The lowest value of the quark mass is dictated by the numerical stability of our algorithm. Thus our calculations of $\vev{j_{\mu}^{f \, 2}}$ at $m_q = 50 \, MeV$ should be considered as only the lower bound for $\vev{j_{\mu}^{f \, 2}}$ at realistic masses of $u$ and $d$ quarks.

\begin{figure}[b]
 \sidecaption
 \includegraphics[width=6cm, angle=-90]{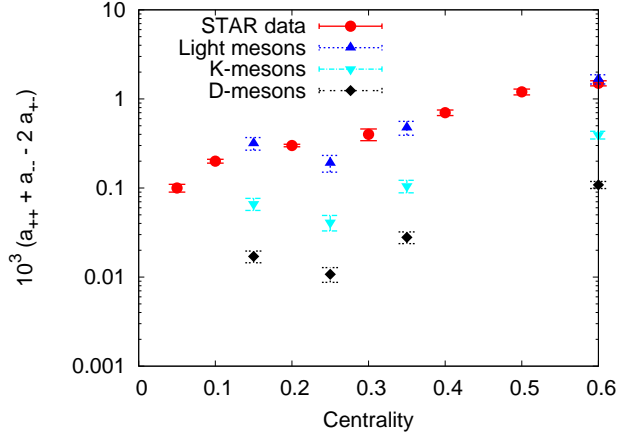}
 \caption{Comparison of the quantity $a_{++} + a_{--} - 2 a_{+-}$ for the experimental data by the STAR collaboration \cite{Abelev:09:1} with the estimates (\ref{dQ_curr2_fin}) based on the results of lattice simulations at different quark masses.}
 \label{fig:cme_cmp}
\end{figure}

 On Fig. \ref{fig:cme_cmp} we compare the quantity $a_{++} + a_{--} - 2 a_{+-}$ calculated for the experimental data by the STAR collaboration \cite{Voloshin:08:1} with the estimate (\ref{dQ_curr2_fin}) based on the results of lattice simulations at different quark masses and different temperatures. In order to match the value of the magnetic field strength in experiment and in simulations, we take the rough estimate from eq. (A.12) in \cite{Kharzeev:08:1}:
\begin{eqnarray}
\label{magn_field_estimate}
e B \sim (0.1 b/R) \, \mbox{GeV}^{2} ,
\end{eqnarray}
where $b$ is the impact parameter. This rough fit also agrees by the order of magnitude with the results of more sophisticated calculations of the magnetic field within the UrQMD model \cite{Skokov:09:1}.

 One can see that the best agreement with the STAR data is obtained at the smallest quark mass. The combination $a_{++} + a_{--} - 2 a_{+-}$ quickly decreases as the quark mass is increased - approximately by a factor of $5$ as the quark mass changes from $50 \, MeV$ to $110 \, MeV$, and by a factor of almost $20$ as the quark mass further increases to $1 \, GeV$. Since all observables $a_{ab}$ are typically of the same order, one can expect that each such observable will also decrease with the quark mass. This dependence of asymmetry of angular distributions of mesons of different flavors on their mass can be used to discriminate between the CME and other phenomena which might cause such asymmetry.

The result may be compared with the perturbative analog of CME \cite{Teryaev:11:1}
resulting from straightforward generalization of Heisenberg-Euler Lagrangian depending on quark mass as $m_q^{-4}$
\begin{eqnarray}
\label{cme}
j_{\mu}^{q}
 =\frac{7 \alpha \alpha_s}{45 m_q^4} \tilde F_{\mu \nu} \partial^\nu (G \tilde G)
\end{eqnarray}
The correspondence of this perturbative and Abelian effect to CME is manifested by the substitution
\begin{eqnarray}
\label{cme_substitution}
\frac{1}{m_q^4} \partial^\nu (G \tilde G)  \to  \partial^\nu \int d^4 z (G \tilde G) \to \partial^\nu \theta
\end{eqnarray}
As a result, the perturbative mechanism may become essential when quark mass is exceeding the inverse correlation length of topological charge density.  One may expect that the transition point from non-perturbative to perturbative mass dependence is not too far from the strange quark mass, like it happen for vacuum quark condensates and strangeness polarization in nucleons (see \cite{Teryaev:11:1} and Ref. therein) so that the non perturbative lattice results are applicable for experimentally important case of strangeness separation.

\section{Induced conductivity and abnormal dilepton yield}
\label{sec:dilepton}

 Another phenomenon which might be caused by superstrong magnetic fields acting on the hadronic matter is the induced anisotropic conductivity along the magnetic field \cite{Buividovich:10:1}. While the Chiral Magnetic Effect is related to the local fluctuations of current density, the induced conductivity reflects the fact that these fluctuations also have long-range correlation in time. Indeed, by virtue of the Green-Kubo relations the conductivity is related to the zero-frequency limit of the spectral function $\rho_{\mu\nu}\lr{w}$ which corresponds to the correlator $\vev{ {\mathcal T} \, j_{\mu}\lr{x} j_{\nu}\lr{y} }$ in Minkowski space \cite{Kadanoff:63:1}:
\begin{eqnarray}
\label{GreenKubo}
\int d^3\vec{x} \vev{ {\mathcal T} \, j_{\mu}\lr{\vec{0}, 0} j_{\nu}\lr{\vec{x}, \tau}}
 =  
\int \limits_{0}^{+\infty} \frac{d w}{2 \pi}\,
\frac{\cosh{ w \, \lr{\tau - \frac{1}{2 T} }}}{\sinh\lr{\frac{w}{2 T}}} \, \rho_{\mu\nu}\lr{w}
\end{eqnarray}
 This spectral function can also be extracted from the results of lattice simulations using the so-called Maximal Entropy Method \cite{Aarts:07:1, Asakawa:01:1}.

 The spectral function $\rho_{\mu\nu}\lr{w}$ determines also the dilepton emission rate from either cold or hot hadronic matter \cite{McLerran:85:1, Teryaev:95:1}:
\begin{eqnarray}
\label{DileptonEmissRate}
\frac{R}{V} = -4 e^4 \int \frac{d^3 p_1}{\lr{2 \pi}^3 2 E_1}
\frac{d^3 p_2}{\lr{2 \pi}^3 2 E_2}
L^{\mu\nu}\lr{p_1, p_2}  \frac{\rho_{\mu\nu}\lr{q}}{q^4},
\end{eqnarray}
where $p_1$ and $p_2$ are the momenta of the leptons, $q = p_1 + p_2$,  $$L^{\mu\nu} = \lr{\lr{p_1 \cdot p_2 + m^2} \eta^{\mu\nu} - p_1^{\mu} p_2^{\nu} - p_2^{\mu} p_1^{\nu}}$$ is the dilepton tensor ($\eta^{\mu\nu}$ is the Minkowski metric), $m$ is the lepton mass. Thus the low-momentum limit of $\rho_{\mu\nu}\lr{w}$ is related, on the one hand, to the emission rate of soft dileptons, and, on the other hand, to the conductivity of hadronic matter.

 The enhancement of the conductivity due to the magnetic field should thus lead to the enhancement of the dilepton emission rate. This might provide a viable explanation of the abnormal soft dilepton yield observed in heavy-ion collisions \cite{Adare:09:1}. Moreover, the anisotropy of the conductivity should lead to specific correlations between the momenta of the dileptons and the direction of the magnetic field (in other words, with the orientation of the reaction plane). 

 In a similar way, one can also consider the correlators of charged vector currents $j_{\mu}^{ff'} = \vev{\bar{q}^f \gamma_{\mu} q^{f'}}$, where $f \neq f'$ are the flavour indices. The correlators of charged vector currents have been studied recently in \cite{Buividovich:11:2} in order to test the conjectured $\rho$-meson condensation in superstrong magnetic fields \cite{Chernodub:10:1}. It was found that as the magnetic field is switched on, this correlator decays slower, which corresponds to the the increase of the corresponding spectral function at lower frequencies. Since charged currents are coupled to leptons and neutrinos, this effect might be manifested in the enhanced lepton and neutrino yield. Excitations over the $\rho$-meson condensate might show up then in the usual decay channel $\rho^{\pm} \rightarrow \pi^{0} \pi^{\pm}$.

 Lattice simulations show that the electric conductivity is nonzero only in the direction of the magnetic field and depends linearly on $q B$ \cite{Buividovich:10:1, Buividovich:10:4}. For sufficiently small momenta  $p_1$ and $p_2$ (and thus for small $q$) by virtue of the Green-Kubo theorem one has $\rho_{ij}\lr{q} \approx \sim \sigma_{ij} q/T \sim B_i B_j q/ \lr{|B| T}$ \cite{Kadanoff:63:1, Aarts:07:1}. Let us also neglect the lepton masses and go to the rest frame of the dilepton pair, where $\vec{p_1} = - \vec{p_2} \equiv p \vec{n}$, $q = \lr{2 p, \vec{0}}$ and the spatial components of the dilepton tensor are $L^{ij} = p^{2} \, \lr{ \delta_{ij} -  n_i \, n_j}$. The dilepton emission rate is therefore proportional to
\begin{eqnarray}
\label{DileptonEmissRateAnisotr}
\frac{R}{V} \sim \int \frac{d^3 p}{\lr{2 \pi}^3 32 E B p^2} \, \lr{\vec{B}^2 - \lr{\vec{B} \cdot \vec{n}}^2}
 \sim  |B| \sin^2\lr{\theta} ,
\end{eqnarray}
where $\theta$ is the angle between the spatial momentum of the outgoing leptons and the magnetic field. Therefore, there should be more soft dileptons emitted perpendicular to the magnetic field than parallel to it. As a result, they are to large extend hidden inside the hadrons in the scattering plane which should lead to the difficulty in their experimental observation. Similar predictions can be also made for the angular distributions of $\pi^{\pm} \pi^{0}$ pairs in the case of charged meson decays (see above).

 In order to estimate the effect of the magnetic field on the total dilepton yield in heavy-ion collisions, we normalize the conductivity induced by the magnetic field to the conductivity at zero magnetic field and at the temperature close to the deconfinement phase transition, $T = 1.12 \, T_c$. By virtue of (\ref{DileptonEmissRate}), the ratio of these conductivities should be equal to the ratio of dilepton emission rates in the low-momentum region. The relevant lattice data is summarized in \cite{Buividovich:10:1, Buividovich:10:4}. In these works it was found that in the deconfinement phase the conductivity is practically independent of the magnetic field. Therefore here we will try to estimate possible contribution of the magnetic field to the dilepton emission rate from hadronic matter in the confinement phase. From the data presented in \cite{Buividovich:10:1, Buividovich:10:4}, we estimate the ratio of the induced conductivity $\sigma\lr{B, T < T_c}$ to the conductivity of quark-gluon plasma $\sigma\lr{B=0, T=1.12 T_c}$ as:
$$
\sigma\lr{B, T < T_c}/\sigma\lr{B=0, T = 1.12 T_c} \approx \frac{e B}{\lr{0.5 \, GeV}^2}
$$
We now use the estimate (\ref{magn_field_estimate}) for the magnetic field strength and take into account that the ratio of conductivities should be equal to the ratio of dilepton emission rates. We thus obtain for the contribution of the magnetic field to the dilepton emission rate:
\begin{eqnarray}
\label{conductivity_ratios1}
R\lr{B, T = 0}/R\lr{B=0, T = 1.12 T_c} \approx \frac{0.4}{3} \, b/R ,
\end{eqnarray}
where the factor $1/3$ appears after averaging the expression (\ref{DileptonEmissRateAnisotr}) over the angle $\theta$ .

 We conclude that at large impact parameters ($b \sim 2 R$) the dilepton yield can increase by up to $20-30 \%$ due to the influence of the magnetic field. This factor should be essentially reduced because the magnetic-induced dileptons are hidden inside the scattering plane. The observed abnormal dilepton yield is, however, maximal for central collisions (where it reaches several hundred percent as compared to the hadron resonance model) and decreases as the impact parameter grows \cite{Adare:09:1}. Such behavior might be caused by several factors, such as the change of the temperature within the fireball or the change of the fireball volume with impact parameter. A proper investigation of such factors is out of the scope of this paper, and cannot be undertaken using the methods of lattice gauge theory.

\section{Conclusions}
\label{sec:conclusions}

 In this paper we have summarized the main experimental signatures of the effects caused by superstrong magnetic fields in heavy ion collisions, namely, the Chiral Magnetic Effect and the abnormal dilepton yield.

 The Chiral Magnetic Effect \cite{Kharzeev:08:1} results in preferential emission of charged hadrons in the direction perpendicular to the reaction plane. The origin of the Chiral Magnetic Effect is the fluctuations of chirality, which are suppressed as the quark mass grows. Thus this asymmetry in angular distributions of charged hadrons, characterized by the coefficients $a_{ab}$ (\ref{observables}), should be strongly suppressed for strange or charmed hadrons.

 The abnormal dilepton yield with specific angular dependence (\ref{DileptonEmissRateAnisotr}) is the consequence of electric conductivity of the hadronic matter induced by the magnetic field. In this case more dileptons are emitted in the direction perpendicular to the magnetic field. Let us also note that since according to our lattice data the magnetic field influences the conductivity only in the confinement phase \cite{Buividovich:10:1}, the significant change of dilepton yield in noncentral heavy-ion collisions might be a signature of the confinement-deconfinement phase transition.

 More generally, lattice data suggests that the influence of the magnetic field on the properties of hadronic matter is stronger in the confinement phase. Thus heavy-ion collision experiments on colliders with lower beam energy but with larger luminosity (such as FAIR in Darmstadt, Germany or NICA in Dubna, Russia) might be more advantageous for studying magnetic phenomena.

\begin{acknowledgement}
 The authors are grateful to M. N. Chernodub, A. S. Gorsky, V.I. Shevchenko, M. Stephanov and A. V. Zayakin for interesting and useful discussions. We are also deeply indebted to D. E. Kharzeev for very valuable and enlightening remarks on the present work. The work was supported by the Russian Ministry of Science and Education under contract No. 07.514.12.4028, by the Grant RFBR-11-02-01227-a of
the Russian Foundation for Basic Research and by the Heisenberg-Landau program of JINR. P.B. was supported by the Sofja Kowalewskaja award from the Alexander von Humboldt Foundation. Numerical calculations were performed at the ITEP computer systems "Graphyn" and "Stakan" (authors are much obliged to A.~V.~Barylov, A.~A.~Golubev, V.~A.~Kolosov, I.~E.~Korolko, M.~M.~Sokolov for the help), the MVS 100K at Moscow Joint Supercomputer Center and at Supercomputing Center of the Moscow State University.
\end{acknowledgement}


\end{document}